\documentstyle[twoside,fleqn,espcrc2,epsfig]{article}

\def\nostrocostrutto#1\over#2{\mathrel{\mathop{\kern 0pt \rlap
  {\raise.2ex\hbox{$#1$}}}
  \lower.9ex\hbox{\kern-.190em $#2$}}}

\newcommand{\be}{\begin{equation}}
\newcommand{\ee}{\end{equation}}
\newcommand{\ba}{\begin{eqnarray}}
\newcommand{\ea}{\end{eqnarray}}

\newcommand{\eref}[1]{(\ref{#1})}      

\newcommand{\N}{{\mathcal N}}

\title{Perturbative description of inclusive energy  spectra}
\author{Sergio Lupia
\address{Max-Planck-Institut f\"ur Physik
(Werner-Heisenberg-Institut),
F\"ohringer Ring 6, D-80805 M\"unchen, Germany.
{\rm E-mail: lupia@mppmu.mpg.de} } }

\begin{document}
\pagestyle{empty}

\null

{\large

\rightline{MPI-PhT/96-72}
\rightline{August 1996}
\vspace{2cm}

 \centerline{\LARGE\bf Perturbative description of inclusive energy
 spectra\footnote{\normalsize
 to be published in the Proceedings of the
 High Energy Conference on Quantum Chromodynamics (QCD '96),
Montpellier, France, July 4$^{th}$-12$^{th}$, 1996,
Ed. S. Narison, Nucl Phys. B (Proc. Suppl.)}  }

\vspace{1.0cm}

\centerline{SERGIO LUPIA}

\vspace{1.0cm}

\centerline{\it Max-Planck-Institut f\"ur Physik}
\centerline{\it (Werner-Heisenberg-Institut)}
\centerline{\it F\"ohringer Ring 6, D-80805 M\"unchen, Germany}
\centerline{E-mail: lupia@mppmu.mpg.de}

\vspace{3.0cm}
\centerline{\bf Abstract}
\bigskip

\noindent
The recent LEP-1.5 data on charged particle inclusive energy
spectra are analyzed within
the analytical QCD approach based on Modified Leading Log Approximation
plus Local Parton Hadron Duality.
The shape, the position of the maximum and the cumulants moments
of the inclusive energy spectrum are well
described within this model. The sensitivity of the results to the running of
the coupling is pointed out.
A scaling law for the one-particle invariant density $E
\frac{dn}{d^3p}$ at small momenta is observed, consistently with the
predictions of colour coherence in  soft gluon bremsstrahlung.

}

\newpage
\begin{abstract}
The recent LEP-1.5 data on charged particle inclusive energy
spectra are analyzed within
the analytical QCD approach based on Modified Leading Log Approximation
plus Local Parton Hadron Duality.
The shape, the position of the maximum and the cumulants moments
of the inclusive energy spectrum are well
described within this model. The sensitivity of the results to the running of
the coupling is pointed out.
A scaling law for the one-particle invariant density $E
\frac{dn}{d^3p}$ at small momenta is observed, consistently with the
predictions of colour coherence in  soft gluon bremsstrahlung.
\end{abstract}
\maketitle
\thispagestyle{empty}


\section{QCD DESCRIPTION OF INCLUSIVE PARTICLE ENERGY DISTRIBUTION}

Analytical predictions of inclusive observables in jet physics can be
performed at parton level
in the framework of the Modified Leading Log Approximation (MLLA)
of QCD\cite{DKMTbook}. In this approach, parton production in jets is described
in terms of a shower evolution which intrinsically  includes coherence, takes
care of both  leading collinear and infrared singularities and of
 energy-momentum conservation (recoil effect).
Predictions depend on two free parameters only,
 i.e., the infrared cut-off at which the parton evolution is stopped, $Q_0$,
 and the effective QCD scale $\Lambda$ which appears in the
 one-loop expression for the running coupling.
To connect predictions at parton level with experimental data,
Local Parton Hadron Duality (LPHD)\cite{LPHD,ko} is taken
as hadronization prescription, i.e., the
inclusive hadron spectra  are required to be
proportional to the corresponding inclusive parton spectra,
obtained by allowing the parton cascade
to evolve down to a cut-off $Q_0$ of the order of few hundreds MeV.
The whole hadronization is then parametrized in terms of only one parameter,
 which gives the overall normalization of the spectrum.

\section{KINEMATICS OF THE SOFT REGION}

The comparison of parton and hadron spectra in the soft region requires some
attention.
On the theoretical side,
 partons are taken as massless and one
does not distinguish between inclusive momentum or energy
spectra\cite{DKMTbook}.
However, a $k_t$ cut-off, $Q_0$, is introduced  for regularization
of the infrared and collinear singularities, and the theoretical spectrum has a
kinematical cut-off at
 $\xi = Y \equiv \log \sqrt{s}/2 Q_0$.
Experimentally, the momentum and energy spectra show
a different behavior at  small momenta due to the effect of hadron mass;
even though this difference does not affect
essentially the gross features of the
spectra, like for instance the peak position,
it is relevant  in the soft region.

To safely connect the two distributions at parton and hadron level,
a prescription has been proposed in \cite{lo}:
the cut-off $Q_0$ has been associated  with an effective hadron
mass\cite{LPHD} such that $E_h = \sqrt{p_h^2+Q^2_0}$
and one has required:
\be
D(\xi, Y) \equiv E \frac{dn(\xi_E)}{dp}  \quad ,  \quad \xi \equiv
\xi_E = \log \frac{\sqrt{s}}{2E}
\label{dual}
\ee
In this way,  a common kinematical behaviour
at hadron and parton levels is indeed obtained.

\section{THE ANALYSIS OF THE SHAPE}

\begin{figure}
          \begin{center}
\mbox{\epsfig{file=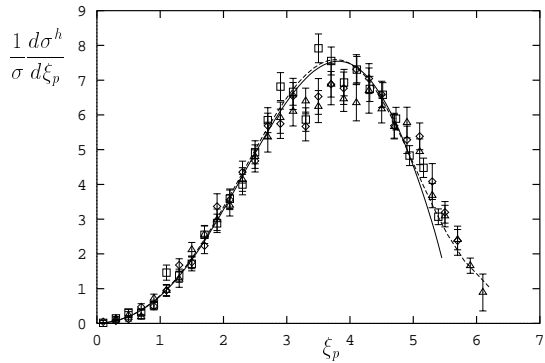,bbllx=3.cm,bblly=18.5cm,bburx=22.5cm,bbury=27.cm,width=10cm}
}
       \end{center}
\caption{Charged particle inclusive momentum distribution at
LEP-1.5 from ALEPH (diamonds),
DELPHI (squares) and OPAL Collaborations
(triangles) in comparison with  theoretical predictions of the
Limiting Spectrum with $Q_0$ = 270 MeV (solid line). Dashed
lines show the predictions of the Limiting Spectrum after  correction for
kinematical effects.}
\label{shape}
\end{figure}

Figure~\eref{shape} shows
experimental data on charged particle inclusive momentum
distribution, $dn/d\xi_p$ as a function of
$\xi_p =  \log (\sqrt{s}/2 p)$,  extracted at LEP-1.5 $cms$ energy  by
 ALEPH, DELPHI and OPAL Collaborations\cite{klo}. Also shown are
the theoretical predictions of the
Limiting Spectrum with
the $Q_0$ parameter taken equal to 270 MeV, as
suggested in ~\cite{lo}, and the free
overall normalization factor fixed to the value  $K^h$ = 1.31.

The Limiting Spectrum predictions with this choice of parameters reproduce
well the  shape around the peak, even though small deviations, compatible
with statistical uncertainties,  are
visible in the region  around the maximum.
The agreement is good in the  large $x_p$ (small $\xi_p$) region as well,
whereas in the small $x_p$ (large $\xi_p$) region,
data show a tail which is not
well reproduced by theoretical predictions.
The tail can be described if one properly takes into account the kinematical
effects discussed in the previous section.
By using the relation~\eref{dual}
 between parton and hadron spectra, the
theoretical predictions for the inclusive momentum spectrum is related to the
Limiting Spectrum prediction $D_q^{\lim}(\xi,Y)$ by the following relation:
 \be
 \frac{1}{\sigma}\frac{d\sigma^h}{d\xi_p} = 2 K^h
 \frac{p}{E} D^{\lim}_q(\xi_E,Y)
 \label{modified}
 \ee
where the factor 2 takes into account that we are considering the full $e^+e^-$
events, whereas $D_q^{\lim}(\xi,Y)$ refers to a single quark jet.

The dashed line in Figure~\eref{shape} shows the prediction of
eq.~\eref{modified} with $K^h$ = 1.34.
 For $E \gg Q_0$ the difference between energy and momentum becomes negligible
  and $dn/d\xi_p \simeq dn/d\xi_E$, $\xi_E \simeq \xi_p$. The good agreement at
  small and intermediate $\xi_p$ is then not modified.
However, this correction becomes important at large $\xi_p$ (small momentum),
where the rescaled prediction
closely  follows the tail of the experimental data.

\begin{figure}
          \begin{center}
\mbox{\epsfig{file=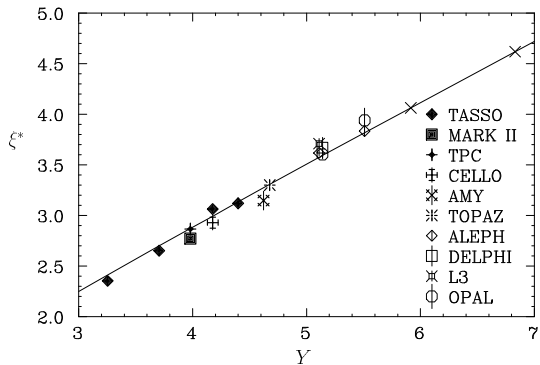,bbllx=2.5cm,bblly=16.5cm,bburx=24.5cm,bbury=24.cm,width=10cm}
}
       \end{center}
\caption{Maximum of the inclusive momentum distribution $\xi^*$
as a function of $Y$;
comparison between experimental data  and theoretical prediction
numerically extracted from the shape of the Limiting Spectrum (solid line);
$Q_0$ = 270 MeV.
Crosses are put at $cms$ energies $\protect\sqrt{s}$ = 200 GeV and 500 GeV.}
\label{maximum}
\end{figure}

\section{THE POSITION OF THE MAXIMUM}

An easily accessable characteristic of the $\xi$-distribution is its maximum
$\xi^*$.  The high energy behaviour of this
quantity is predicted in MLLA for the Limiting Spectrum as~\cite{DKMTbook}:
\begin{equation}
\xi^*_{asy} \; = \; Y \: \left [ \frac{1}{2} \: + \: \sqrt{
\frac{C}{Y}} \: - \: \frac{C}{Y} \right ]
\label{asy}
\end{equation}
\noindent with the constant term  given by
$C  =  \frac{a^2}{16 \: N_C b}  =  0.2915 (0.3513)$
for $n_f  =  3 (5)$.
It is actually possible to extract  numerically the position of the
maximum $\xi^*$ from the shape of the Limiting Spectrum itself;
even though the asymptotic formula describes very well the energy dependence
of the actual maximum, it underestimates the actual value by an (approximately)
constant value  of the order of
0.1 units. This gives rise to a difference of the  order of 20-30 MeV
in the determination of  the best value of the cut-off $Q_0$  from the energy
dependence of the maximum.

In Figure~\eref{maximum} we compare experimental data on
the maximum $\xi^*$ extracted from Gaussian or Distorted Gaussian
fits to the central region of the inclusive momentum
spectrum with the theoretical
predictions obtained extracting the maximum from the shape of the Limiting
Spectrum\cite{klo}.
A good agreement is found for $Q_0$ = 270 MeV and $n_f$ = 3.
Notice the choice of 3 active flavours, as
suggested by the study of moments in \cite{lo},
contrary to the usual approach, in which
5 active flavours are used.
In this respect, let us remind
that the number of flavours enters in the expressions for the cumulant moments
through the running coupling $\alpha_s(y,n_f)$; since the MLLA is defined at
one-loop level, there remains a scale ambiguity in the
 expression for $\alpha_s$ and
 it is then not so straightforward to fix the scale and
heavy quark thresholds. As discussed in more detail in \cite{klo},
 kinematical reasons suggest to push the
thresholds to larger scales and keep 3 active flavours only.
In any case, for the maximum $\xi^*$ the effect of different choices on the
number of active flavours and on the treatment of heavy quark thresholds
 is of the order of 1\%  only.

\section{MOMENT ANALYSIS}

\begin{figure}
          \begin{center}
\mbox{\epsfig{file=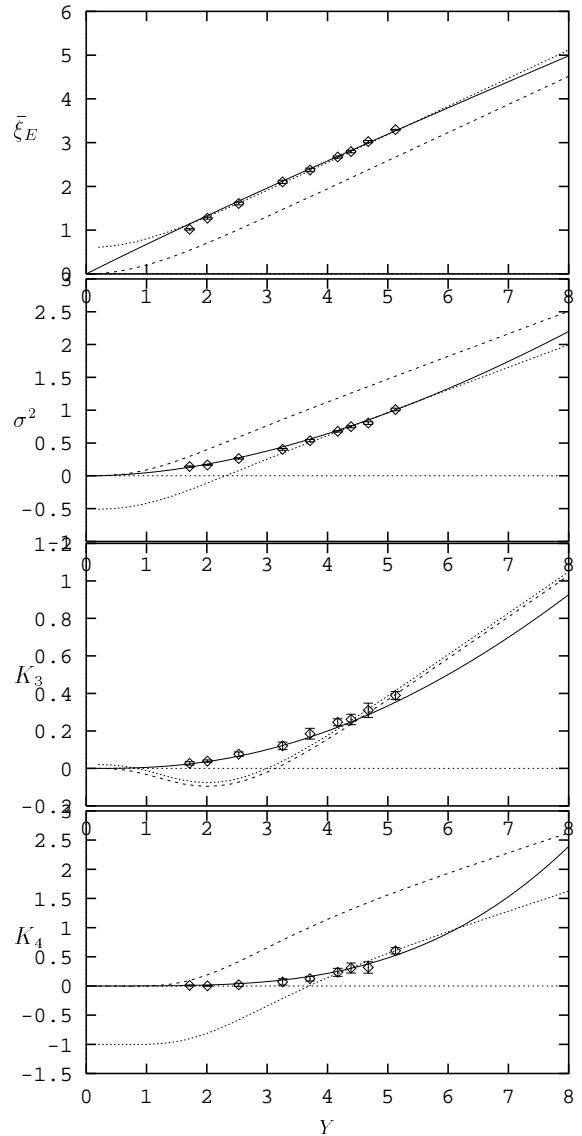,bbllx=4.5cm,bblly=5.cm,bburx=20.cm,bbury=26.cm,height=14.cm}
}
       \end{center}
\caption{The first four
cumulants of charged particles' energy spectra $E dn/dp$ vs. $\xi_E$,
are shown as a function of $Y$
for $Q_0$ = 270 MeV.
Predictions of the Limiting Spectrum of MLLA with running $\alpha_s$ (solid
lines), of MLLA with fixed $\alpha_s$ (= 0.214) (dashed lines) and of MLLA
with fixed $\alpha_s$ normalized by hand to the experimental point
at $\protect\sqrt{s}$ = 44 GeV (dotted lines) are also shown (in all cases
$n_f$ = 3).}
\label{moments}
\end{figure}

In order to study the inclusive spectrum in more detail, it is convenient to
look at the energy dependence of its moments
or cumulant moments\cite{lo,DKTInt}.
Note that cumulants of order $q \ge$ 1
are independent of the overall normalization, which is the main source of
systematic uncertainties; in addition, theoretical
 predictions for these observables are absolutely normalized at threshold and
then  allow  to test the perturbative picture down to low $cms$ energies.

The moments $< \xi^q>$ and the cumulants $K_q$
can be extracted  from the spectrum $E dn/dp$ as a function of  $\xi_E$:
\be
< \xi^q> \equiv \frac{1}{\bar N_E} \int d\xi \xi^q D(\xi) \; .
\ee
The average multiplicity $\bar \N_E$ is obtained as the
integral over $\xi_E$ of the full spectrum $E dn/dp$ (zero-th order moment).
For the unmeasured interval near $\xi_E \simeq Y$
(small momenta) a contribution is found by linear extrapolation as in
\cite{lo}.

The cumulants up to order $q$=4
are shown in Figure~\eref{moments} as a function of $Y$.
Theoretical predictions  for the Limiting Spectrum with
$Q_0$ = 270 MeV\cite{DKTInt} are also shown (solid line).
The agreement with experimental results  is very satisfactory.
The dashed lines show the predictions of the MLLA, but
with fixed coupling $\alpha_s$; this model cannot reproduce the behavior of
high order cumulants, showing the sensitivity of the moments of the inclusive
spectra to the running of the coupling.

 It is possible to relax
the absolute normalization, thus building an effective model with
one more parameter for each cumulant. By rescaling the fixed $\alpha_s$ model
to reproduce the high energy region (in particular we fixed the curve at the
data point at $\sqrt{s}$ = 44 GeV),  one  can
describe the experimental data only in a small  energy region
 (about one unit in $Y$) (dotted lines). Also in this case,
the predictions differ from the full
MLLA model with running $\alpha_s$ at larger $cms$ energies; in this respect,
the study of cumulant moments of the inclusive energy spectra
at $p\bar p$ collisions, where larger values of $Y$ can be
reached\cite{korytov}, could give new and very interesting information.

\section{THE BEHAVIOR OF THE INVARIANT CROSS SECTION IN THE SOFT LIMIT}

\begin{figure}
          \begin{center}
   \mbox{
\mbox{\epsfig{file=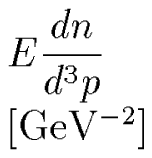,bbllx=5.2cm,bblly=17.5cm,bburx=5.4cm,bbury=25.cm}}
\mbox{\epsfig{file=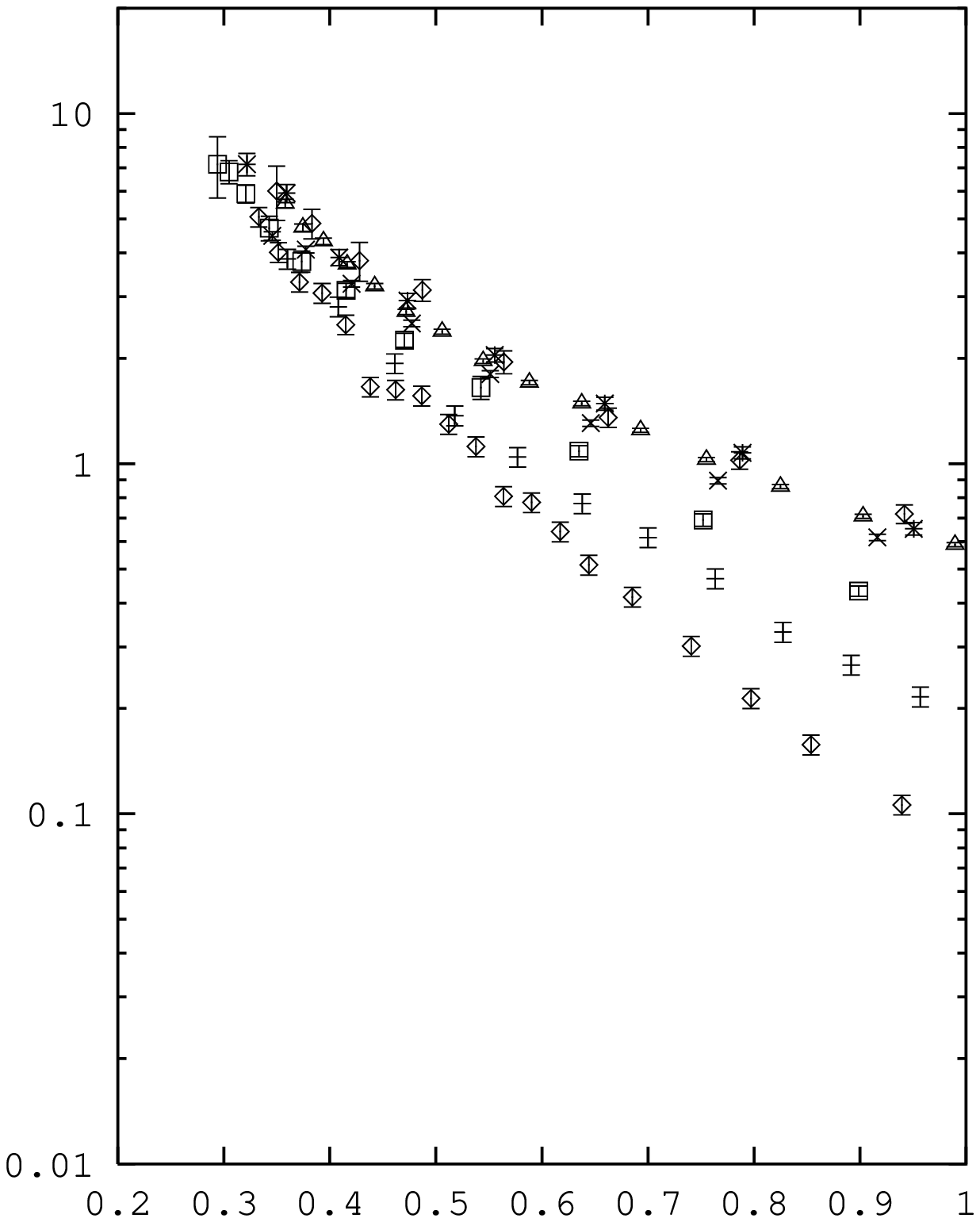,width=.72\linewidth,bbllx=5.5cm,bblly=2.5cm,bburx=15.cm,bbury=18.cm}}
  }        \end{center}
\vspace{-0.3cm}
\centerline{$\quad \qquad E$ [GeV]}
\caption{Invariant distribution $E dn/d^3p$ for charged particle
as a function of the particle energy $E$ with $Q_0$ = 270 MeV.}
\label{charged}
\end{figure}

\begin{figure}
          \begin{center}
   \mbox{
\mbox{\epsfig{file=scritte.ps,bbllx=5.2cm,bblly=17.5cm,bburx=5.4cm,bbury=25.cm}}
\mbox{\epsfig{file=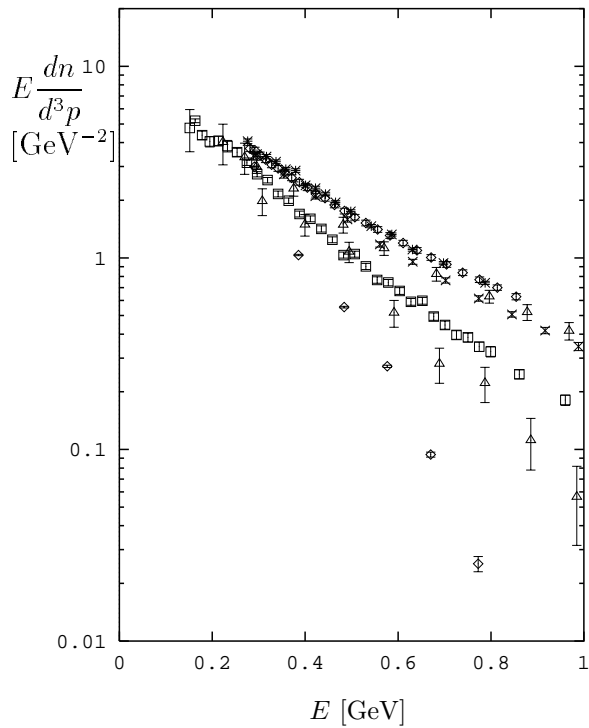,width=.72\linewidth,bbllx=5.5cm,bblly=2.5cm,bburx=15.cm,bbury=18.cm}}
  }        \end{center}
\vspace{-0.3cm}
\centerline{$\quad \qquad E$ [GeV]}
\caption{Same as in~\protect\eref{charged} but for charged pions with
$Q_0$ = $m_{\pi}$.}
\label{pion}
\end{figure}


Figure~\eref{charged}  shows
the charged particle  invariant cross section, $E dn/d^3p$,
as a function of the particle energy $E$ at different $cms$ energies
ranging from 3 GeV up to LEP-1.5\cite{lo,klonew}.
The value of 270 MeV has been used for the effective mass  $Q_0$
which enters in the kinematical relation, $E^2 = p^2 + Q_0^2$.
It is remarkable that
all curves scale  with $cms$ energies within 20\%
at particle energy of the order of few hundreds MeV;
at larger particle energies, a violation of the scaling-law is visible.
In addition, all curves  approach a finite limit.
LEP data seems to tend to a larger limiting value; it is not clear whether this
is due to a different physics at LEP energies or rather an
overall  systematic effect in the normalization of the different experiments.

These results are particularly interesting, since colour coherence
predicts indeed that soft particles do not multiply\cite{LPHD}; the invariant
cross-section should then  be independent of $cms$ energy
at low particle energy and  should approach a finite limit\cite{klonew}.
The experimental observation of this scaling law suggests
that the invariant cross-section keeps trace of the effects of colour
coherence present at parton level.


We have considered so far the inclusive spectra for all charged particles only.
It is interesting to study whether the perturbative approach can be extended
to identified particle spectra, in particular $\pi$, $K$ and $p$ spectra.
In this case, a natural mass scale would be provided by
 the mass of the particle itself.
The success of this approach with the identification of the particle mass as
 the theoretical cut-off $Q_0$ for describing the hump-backed plateau
  is still controversial  (see \cite{ko} for a discussion of this point).
Here we concentrate on the behaviour of the
invariant cross-section in the soft region.
Figure~\eref{pion} shows the invariant cross section $E dn/d^3p$ for charged
pions
as a function of the particle energy $E$
extracted from inclusive momentum spectra at $cms$ energies ranging from
1.6 GeV up to LEP-1 energy.
Also in this case an approximate scaling law with $cms$ energy
is seen  at particle energy of few hundreds
MeV and a finite limit is approached.
Work is in progress\cite{klonew} to further investigate the consequences of
this result and to extend this analysis to other particles' species.

\section{CONCLUSIONS}

The analysis of the recent LEP-1.5 data shows that the analytical
perturbative description of inclusive particle distributions based on MLLA plus
LPHD is in a good shape.  The charged
particle spectrum (up to the overall normalization), the position of its
maximum and the cumulants moments  can be described by the Limiting Spectrum
with only one adjustable scale parameter $Q_0$ = $\Lambda$.
 No additional sizeable effects from hadronization are
visible for these observables.
The sensitivity of the inclusive particle spectra to the running of the
coupling has been pointed out.
The approach to an energy independent limit of the invariant density $E
dn/d^3p$ for momenta $p \to$ 0 in a wide range of primary
energies $E_{jet}$ in agreement with analytical results from the QCD jet
evolution which includes the coherence of the soft gluon
bremsstrahlung has also been observed.

\section*{ACKNOWLEDGEMENTS}

I thank Valery A. Khoze and Wolfgang Ochs for discussions
and collaboration on the subjects of this talk.
I thank S. Narison for the nice atmosphere created at the
Conference.

\end{document}